\documentclass[10pt]{article}

\usepackage{a4wide,epsfig}

\renewcommand{\thefootnote}{\fnsymbol{footnote}}
\begin{document}    

\title{\vskip-3cm{\baselineskip14pt
\centerline{\normalsize\hfill TTP99--22}
\centerline{\normalsize\hfill hep-ph/9905298}
\centerline{\normalsize\hfill May 1999}
}
\vskip.5cm
{\bf
Automatic application of successive asymptotic expansions of Feynman
diagrams
}
\vskip.3cm
}
\author{T. Seidensticker \\[2em]  
 {\it Institut f\"ur Theoretische Teilchenphysik,}\\
  {\it Universit\"at Karlsruhe, D-76128 Karlsruhe, Germany}
}
\date{}
\maketitle

\begin{abstract}
\noindent
We discuss the program EXP used to automate the successive application 
of asymptotic expansions to Feynman diagrams. We focus on the generation of 
the relevant subgraphs and the determination of the topologies for the 
remaining integrals. Both tasks can be solved by using backtracking-type 
recursive algorithms. In addition, an application of EXP is presented,
where the integrals were calculated using the FORM packages MINCER and MATAD.
\end{abstract}

\thispagestyle{empty}
\setcounter{page}{1}

\renewcommand{\thefootnote}{\arabic{footnote}}
\setcounter{footnote}{0}

\section{Introduction}

The necessity of calculating higher order corrections to physical processes
is based on the increasing precision of experimental data.
Unfortunately, analytic expressions for higher order corrections are in
most cases out of reach. This is mainly due to the fact that the integrals
contain several different mass scales. 

One possible approach to tackle the problem is based on the procedures of
asymptotic expansions (see, e.g.~\cite{RefAsyExp} and references therein)
which provide rules for consistent expansions of Feynman diagrams in large 
scales. If a certain hierarchy between the different mass scales can be 
found for a given diagram one expands with respect to a small quantity. 
Although the complexitity of the remaining integrals is reduced by the 
asymptotic expansion sometimes those integrals cannot be solved 
analytically either. To further proceed in these cases we repeat the 
application of an expansion procedure using another large scale. 
Consequently the final result is expressed in terms of nested series in small
quantities.     

In this paper we discuss algorithms we used to automate the successive
application of asymptotic expansions. The algorithms were implemented in
the C++ program EXP. The output files of EXP can be directly used by a skeleton 
FORM \cite{FORM} {\em mainfile}. The mainfile is based on the integration 
packages MINCER \cite{MINCER} and MATAD \cite{MATAD} which perform the
actual integrations for single scale integrals up to three loops. Also some 
administrative files like make-files are written by EXP.

\section{Asymptotic expansions} 
Let us begin by listing the rules of the asymptotic expansions which are 
expressed in a purely diagrammatic manner. They can be divided into two parts:
First, the rules for the selection of all relevant so-called 
{\em hard subgraphs} and second, how to expand the propagators of the hard 
subgraphs. 

In case of expansions with respect to a large mass (the so-called {\em hard 
mass procedure}) one has to find all hard subgraphs which contain all lines 
carrying the large mass and are one-particle-irreducible with respect to 
light lines in their connected parts. The propagators appearing in the hard
subgraph have to be expanded with respect to small masses and external momenta.

Expansions with respect to a large momentum ({\em large momentum procedure}) 
can be obtained in a similar way. The hard subgraphs must contain the vertices 
with the large external momentum and become one-particle-irreducible if those
vertices are identified. The propagators have to be expanded with respect to
masses and external momenta generated by removing lines from the initial 
diagram.

The final result for a given diagram is obtained in four steps: 

(1) Shrink the lines of the hard subgraph to a point. The remaining diagram 
is called {\em co-subgraph}. 

(2) Expand the propagators and (if necessary) do the integrals in the hard
subgraph. Insert the result into the co-subgraph. 

(3) Calculate the remaining integrals in the co-subgraph.

(4) Sum over all terms. 

The number of relevant subdiagrams increases rapidly with the number of
loops of the initial diagram especially when we deal with successive 
asymptotic expansions. Providing all necessary information by hand is nearly 
impossible and automation is an obvious task.  

\begin{figure}[h!]
\begin{center}
\epsfig{bbllx=145,bblly=637,bburx=455,bbury=730,scale=0.7,file=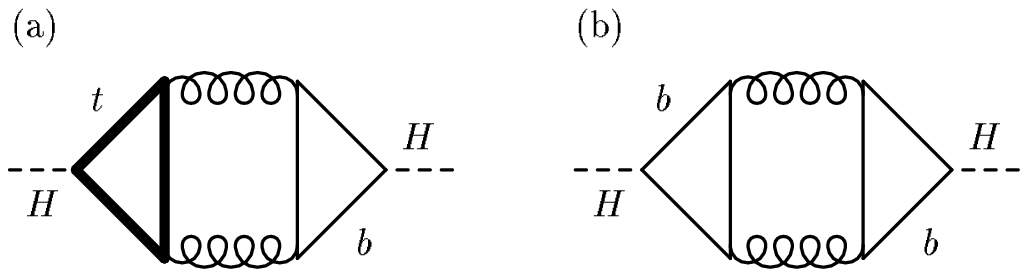}
\end{center}
\caption{\label{fig1}Double triangle graphs contributing to 
         $\Gamma(H \rightarrow b\bar{b})$.}
\end{figure}

\section{\label{sec:example}Example} 

To illustrate the successive application of asymptotic expansions we
consider the singlet contributions of ${\cal O}(\alpha_s^2)$ to the decay of
a scalar Higgs boson into bottom quarks. Instead of calculating the vertex
diagrams, we use the optical theorem and compute the imaginary part of the
Higgs boson propagator to three loops. There are four diagrams relevant for 
this problem, two are shown in figure \ref{fig1}. The dashed lines 
denote the Higgs boson of mass $M_H$, the thick lines represent top quarks 
($M_t$), the thin lines bottom quarks ($M_b$) and the curly lines are gluons. 
The remaining two diagrams have crossed gluon lines and their contribution can
be included by an additional factor of 2 in the result. This problem has been 
analyzed in \cite{CheKwi95,LarRitVer95}. We used the results given in 
\cite{CheKwi95} as a check for the program EXP.  

Suppose we want to compute the singlet contribution for a Higgs boson mass
smaller than the mass of the top quark. Then we have to choose the hierarchy
\begin{displaymath}
M_t^2 \gg q^2 = M_H^2 \gg M_b^2,
\end{displaymath}
where $q$ denotes the external momentum. Therefore we have to compute an
expansion with respect to a large mass ($M_t^2$) followed by an expansion 
with respect to a large momentum ($q^2 = M_H^2$). 

\begin{figure}[t!]
\begin{center}
\epsfig{bbllx=33,bblly=204,bburx=482,bbury=654,scale=0.55,file=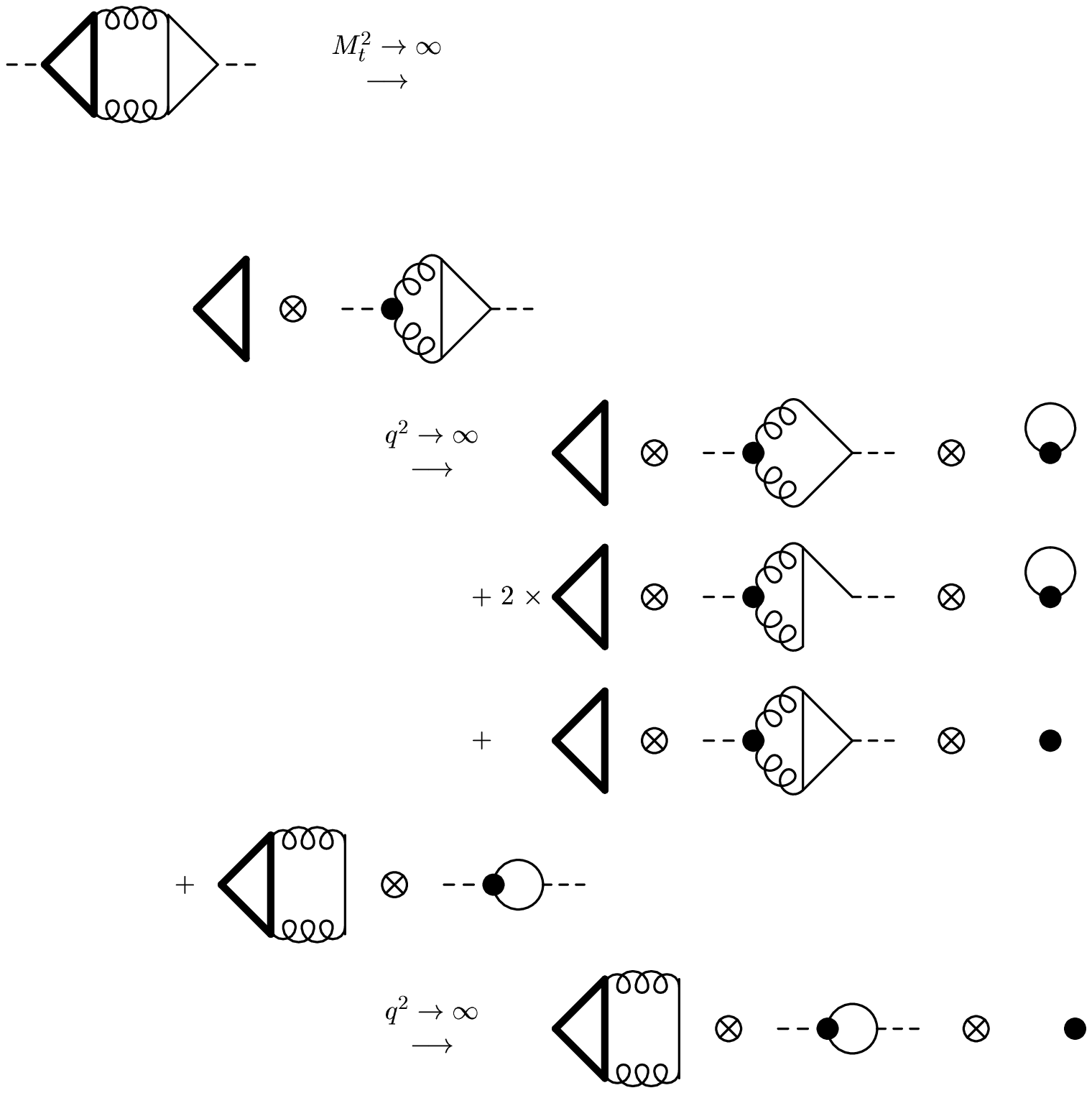}
\end{center}
\caption{\label{fig2}Successive asymptotic expansion of diagram \ref{fig1}(a).}
\end{figure}

If we restrict ourselves to the imaginary part of the Higgs boson propagator,
the hard subgraphs for diagram \ref{fig1}(a) are shown in figure
\ref{fig2}. The application of the hard mass procedure with respect to $M_t$
results in two hard subgraphs shown on the left hand side of figure \ref{fig2}.

In the upper case, the original integral factorizes into a 1-loop tadpole
(hard subgraph) and a 2-scale 2-loop integral (co-subgraph) which has to be
expanded in the next step with respect to the large momentum $q^2$. Note
that there is a contribution from the naive Taylor expansion with respect
to the bottom quark mass $M_b$ (last line of upper case). In the lower case,
the result is a product of a 2-loop tadpole and a 2-scale 1-loop integral.
The 2-scale integral can be solved analytically. Since we use the packages 
MINCER and MATAD -- dealing with single scale integrals -- the 
co-subgraph has to be expanded as well.

\section{Generation of subgraphs}

In this section we will shortly discuss the algorithms used for the
generation of the relevant subgraphs. The main data structure used in the
implementation of EXP is a graph represented by a set of lines which are 
in turn represented as connections of vertices. 

We use a prototype backtracking algorithm to generate all possible subsets of 
lines. For each subset the conditions for hard subgraphs given by the current
procedure are tested. Since these rules are mainly based on the property
``one-particle-irreducible'' we must develop a general prescription for
this to construct an algorithm.  

A graph is one-particle-irreducible if cutting an arbitrary line does
not split the graph into two distinct parts. Therefore every vertex
should still be connected to all the other vertices after removing one line. 
That means that one can find a route through the graph given by a sequence 
of lines to get from one vertex to the other. The procedure of finding
such a route between two vertices can be realised quite elegantly using a 
recursive algorithm: One starts at a given vertex and determines all vertices
directly (by one line) connected to it. Then one recursively tries to 
find a connection between the directly connected vertices and the target 
vertex. Thus all possible routes through the graph are taken into account. 

Further limitations implied by the rules of asymptotic expansion can be
easily implemented since they restrict only the type of lines to be cut or
the subset of vertices to be connected after cutting a line.

\section{Determination of topologies}

Our approach for the determination of the topologies is based on the 
observation that the combinations of external and loop momenta in the lines
of the graph can be used to identify the topology. The idea is to compare a 
given distribution of momenta to all possible ones for a certain number of
{\em standard topologies} which can (or must) be defined by the user. 

The determination of a topology can be divided into four parts: 

(1) A momentum distribution is computed for each individual subgraph. Note 
that not all possible choices are suitable for our problem since the 
propagators must be expanded in taylor series.

(2) The rules for the expansion of the propagators are applied. This is done 
by setting marks on momenta and masses which have to be expanded. Marked 
quantities will be ignored during the next steps, because they are obviously
irrelevant for the determination of topologies. The real expansion is, of
course, done later using FORM. Lines sharing the same loop momenta
are sorted into groups which correspond to different types of integrals. 

(3) All possible distributions of momenta for the standard topologies are
computed and the coefficients of the external and loop momenta are compared to
those calculated for the lines in the current subset. This comparison of the 
external and loop momenta fixes the topologies with respect to the number of
loops and external momenta. 

(4) As the last step, the masses of the propagators will be compared to those 
defined in the standard topology.

The computation of all possible distributions of momenta which is the
crucial part of the algorithm described above is done by recursively
assigning loop momenta to certain lines, i.e.~by defining that certain lines
should carry the loop momenta and solving the linear set of equations given 
by the conservation of momentum at each vertex. Therefore we restrict
ourselves to distributions of momenta with integer coefficients for external
and loop momenta which is sufficient for our problem. 

There will be cases where the algorithm fails to find a topology for a group 
of lines. These lines will be separated from the full graph and an asymptotic
expansion is applied again, until all integrals can be expressed in terms of 
standard topologies. 

The hierarchy of mass scales for all expansions must be defined by the user.

In principle, this algorithm is not limited with respect to the number of
loops or external momenta as long as the user supplies a sufficient amount 
of standard topologies.

\section{Example: result}

In this section we list the result for the imaginary part of the diagrams
considered in section \ref{sec:example} expressed in terms of a nested
series in the the variables $M_H^2/M_t^2$ and $M_b^2/M_H^2$. It reads: 
\begin{eqnarray}
\lefteqn{\Gamma^{sing}(H \rightarrow b\bar{b}, gluons) 
   =  {{N_C \, M_H \, \alpha} \over s_W^2} \,{M_b^2 \over M_W^2} \, 
     \bigg( {\alpha_s \over \pi} \bigg)^2 \, C_F T \times }
   \nonumber\\&&\mbox{}
   \bigg\{ 
        {7 \over 12} + {1 \over 8} \, L_{tH} 
        + {M_b^2 \over M_H^2} \, \bigg( 
          - {3 \over 8} - {3 \over 4} \, L_{tH} - {1 \over 4} \, L_{Hb} 
          \bigg) 
        + {M_H^2 \over M_t^2} \, \bigg(
          {2011 \over 129600} + {41 \over 8640} \, L_{tH} 
          \bigg) 
        \nonumber\\&&\mbox{}
        + {M_b^2 \over M_t^2} \bigg( 
          {713 \over 14400} + {7 \over 1440} \, L_{tH} 
          + {1 \over 288} \, L_{Hb} \bigg) 
        + \bigg( {M_H^2 \over M_t^2} \bigg)^2 \, \bigg(
          {28307 \over 25401600} + {47 \over 120960} \, L_{tH} 
          \bigg)
        \nonumber\\&&\mbox{}
        + {{M_b^2 \, M_H^2} \over M_t^4} \bigg( 
          {10901 \over 25401600} + {1 \over 1080} \, L_{tH}
          \bigg) 
        + {\cal O} (M_b^4)
        + {\cal O} (M_t^{-6})
   \bigg\}, \label{eq:result} 
\end{eqnarray}
with $L_{tH} = \ln(M_t^2/M_H^2)$ and $L_{Hb} = \ln(M_H^2/M_b^2)$. $N_C$
denotes the number of colours and $s_W$ is the sine of the weak mixing
angle; $T = 1/2$ and $C_F = (N_C^2 - 1)/(2 N_C)$. Using EXP we were able to
add the terms proportional to $M_t^{-4}$ to the result given in 
\cite{CheKwi95}. Numerically they are small compared to the leading terms.

The notation $\Gamma^{sing}(H \rightarrow b\bar{b}, gluons)$ has been
chosen, because the diagrams shown in figure \ref{fig1} contain contributions
from cuts through the two gluon lines. These have to be subtracted if one is 
not interested in the total decay rate and are known analytically to this
order \cite{EllGaiNan76}.

There are two important checks for the result: First, it must be independent
of the choice of the QCD gauge parameter. Second, since the diagrams shown 
in figure \ref{fig1} contain no divergent subgraphs their imaginary part 
must be finite without further renormalization.

\section*{Acknowlegements}

We would like to thank K.G. Chetyrkin, R. Harlander, J.H. K\"uhn and 
M. Steinhauser for fruitful discussions and advice, and D. Fliegner for
carefully reading the manuscript. This work was supported by the 
{\it DFG-Forschergruppe "Quantenfeldtheorie, Computeralgebra und 
Monte-Carlo-Simulationen"}, the {\it Landesgraduiertenf\"orderung} and the 
{\it Graduiertenkolleg "Elementarteilchenphysik"} 
at the University of Karlsruhe.

\end{document}